\documentclass[preprint,aps,showpacs,showkeys,nofootinbib]{revtex4-1}
\usepackage{amsmath}
\usepackage{amsfonts,amssymb}
\usepackage{multirow}
\usepackage{graphicx,booktabs,rotating}
\usepackage{float}
\usepackage{appendix}
\usepackage{color}
\usepackage{url}
\usepackage{subfigure}
\usepackage{footnote}

\def\nl{\nonumber\\}

\begin{document}

\title{Semileptonic $B$ and $B_s$ decays involving scalar and axial-vector mesons }
\author{Xian-Wei Kang $^1$}
\author{Tao Luo $^2$} 
\author{Yi Zhang $^2$} 
\author{Ling-Yun Dai $^{3, 4}$}
\author{Chao Wang $^5$}
\affiliation{$^1$ College of Nuclear Science and Technology,
Beijing Normal University, Beijing 100875, China\\
$^2$ Key Laboratory of Nuclear Physics and Ion-beam Application
(MOE) and Institute of Modern Physics, Fudan University, Shanghai 200443, China\\
$^3$ School of Physics and Electronics, Hunan University, Changsha 410082, China\\
$^4$ Institute for Advanced Simulation, Institut f\"ur Kernphysik
and J\"ulich Center for Hadron Physics, Forschungszentrum J\"ulich,
D-52425 J\"ulich, German\\
$^5$ Center for Ecological and Environmental Sciences, Key
Laboratory for Space Bioscience and Biotechnology, Northwestern
Polytechnical University, Xi'an 710072, China}

\begin{abstract}
We report our theoretical calculations on the branching fractions
for the semileptonic $B$ and $B_s$ decays based on the form factors
in the covariant light-front quark model. That is, $B (B_s) \to
(P,\, V,\, S,\,A) \ell \nu_\ell$, where $P$ and $V$ denote the
pseudoscalar and vector mesons, respectively, while $S$ denotes the
scalar meson with mass above 1 GeV and $A$ the axial-vector meson.
The branching fractions for the semileptonic $B\to P$ and $V$ modes
have been measured very well in experiment and our theoretical
values are in good agreement with them. The ones for $B\to S$ and
$A$ modes are our theoretical predictions. There is little
experimental information on the semileptonic $B_s$ decays although
much theoretical effort has been done. In addition, we predict the
branching fractions of $B\to D^*_0(2400) \ell \bar\nu_\ell$ and
$B_s\to D^{*-}_{s0}(2317) \ell \bar\nu_\ell$ as $(2.31\pm
0.25)\times 10^{-3}$ and $(3.07\pm0.34)\times 10^{-3}$, in order,
assuming them as the conventional mesons with quark-antiquark
configuration. The high luminosity $e^+e^-$ collider
SuperKEKB/Belle-II is running, with the data sample enhanced by a
factor of 40 compared to Belle, which will provide huge opportunity
for the test of the theoretical predictions and further help
understand the inner structure of these scalar and axial-vector
mesons, e.g., the glueball content of $f_0(1710)$ and the mixing
angles for the axial-vector mesons. These decay channels can also be
accessed by the LHCb experiment.
\end{abstract}

\maketitle

\section{Introduction}
The CP violation is one of the necessary Sakharov conditions for the
emergence of matter-antimatter asymmetry \cite{Sakharov}, which is a
key question in the nature. The Cabibbo-Kobayashi-Maskawa (CKM)
matrix \cite{CKM} has been an indispensable skeleton of the Standard
Model (SM), which successfully describes the CP violation in the
quark sector. In the CKM matrix, the unitarity relation
$|V_{ub}|^2+|V_{cb}|^2+|V_{tb}|^2=1$ is fulfilled, and any deviation
of this unitarity constraint will be a signal of New Physics. Thus
the precise determination of the CKM matrix elements has been a key
task and activity in the community of flavor physics
\cite{PDG,HFAG}. A recent review for the leptonic and semileptonic
$B$ decays is compiled in Ref.~\cite{Dingfelder}. In our current
work, we will consider $B (B_s) \to (P, V, S, A) \ell \nu_\ell$
decay \footnote{We will consider $B^+$ as an example, and $B^0$
decay is the same as $B^+$ case due to the isospin symmetry. This
point can be verified by the branching fractions $\mathcal{B}(B^+\to
\bar D^0 \ell\nu_\ell) =\mathcal{B}(B^0\to
D^-\ell\nu_\ell)=(2.20\pm0,10)\%$ \cite{PDG}, and
$\mathcal{B}(B^+\to \pi^0\ell\nu_\ell)\approx \mathcal{B}(B^0\to
\pi^-\ell\nu_\ell)/2$, $\mathcal{B}(B^+\to
\rho^0\ell\nu_\ell)\approx\mathcal{B}(B^0\to \rho^-\ell\nu_\ell)/2$
\cite{PDG} due to the additional factor of $1/\sqrt{2}$ appearing in
the quark components in the neutral $\pi$ and $\rho$. We also note
that these channels could also be treated in the PQCD or light-cone
models, e.g., in Refs.~\cite{WeiWang,ZhigangWang,ZhenjunXiao}.},
where $P$ and $V$ denote the pseudoscalar and vector mesons,
respectively, while $S$ denotes the scalar meson with mass above 1
GeV and $A$ the axial vector meson. In these processes, either
$|V_{ub}|$ or $|V_{cb}|$ is involved. And notably, there is a
mismatch for the extraction of $|V_{ub}|$ from the inclusive and
exclusive decays, which is the so-called $|V_{ub}|$ puzzle
\cite{PDG,KangBl4}. The measurement of these channels can certainly
help for rendering more information on the determination of CKM
matrix elements, at least as a supplement to the conventional
exclusive decays. Inversely, to calculate the branching fractions,
we rely on their concrete values from Particle Data Group (PDG)
\cite{PDG}: $|V_{cb}|=(42.2\pm0.8)\times 10^{-3}$, and
$|V_{ub}|=(3.94\pm 0.36)\times 10^{-3}$ as a combination of the
determinations from inclusive and exclusive decays.

For the axial-vector mesons which contain the superposition of quark
contents $s\bar s$ and $q\bar q\equiv (u\bar u + d \bar
d)/\sqrt{2}$, $f_1(1285)$ and $f_1(1420)$, $h_1(1170)$ and
$h_1(1380)$, and $K_{1A}$ and $K_{1B}$ do mix. The mixing angles are
not fully fixed yet, see e.g., Ref.~\cite{ChengMixingAngle}. The
structure of the scalar meson is more obscure, see the review
\cite{PDGreview}. For example, $f_0$ states
($f_0(1370),\,f_0(1500),\,f_0(1710)$) are interpreted as the mixed
states of $q\bar q$, $s\bar s$ and glueball ($G$), but which one
consists mainly of $G$ is not fully determined \footnote{ We also
note that the knowledge of the two-photon couplings to the scalars
is helpful to understand their structures \cite{two-photon}.}. The
observables depend on, or even are sensitive to these mixing angles.
The three-body semileptonic decay is an ideal place to study the
weak hadronic transition form factor as well as the underlying
structure of such mesons due to the absence of the final-state
interactions (FSIs) between hadrons \footnote{The proton-antiproton
FSI has been elaborately examined in various decay channels
\cite{Kangppbar,Daippbar}.}. As such, the theoretical prediction of
the relevant semileptonic decay channels becomes crucial for the
future experimental measurement.

From the experimental point of view, Belle has accumulated huge data
samples, that can be exploited to measure the branching fractions
and hadronic transition form factors for the various semileptonic
decay channels, in order to test or constrain the various
theoretical models. There are $(772\pm11)\times 10^6$ $B\bar B$
\cite{Belle-BBbar} and $(6.53\pm0.66)\times 10^6$ $B_s\bar B_s$
pairs \cite{Belle-BsBsbar} collected at $\Upsilon(4S)$ and
$\Upsilon(5S)$ resonances, respectively, by the Belle detector at
the KEKB asymmetric energy electron-positron collider. The
statistics will be enhanced by a factor of 40 for Belle-II, and by
the mid of next decade, 50 times more data is expected comparing to
the Belle experiment. Our predicted branching fractions are
typically in the order of $10^{-5}$ so that they can be, in
principle, easily accessed by the Belle/Belle-II and LHCb
experiments.

\section{Theoretical framework}
The transition form factor is a probe to the inner structure of the
hadron. Among the various theoretical tools for the form factor, we
will concentrate on the application of the light-front quark model
(LFQM) \cite{Jaus2,Jaus3} and its covariant extension (CLFQM)
\cite{Jaus1}, see also \cite{Ji,Bakker1,Bakker2}. A distinct feature
of the light-front frame is that the diagrams involving quarks
created out of or annihilating into the vacuum can be eliminated,
i.e., only the valence quarks are considered in the meson or baryon
\cite{Brodsky}. This leads to a relativistic quark model which
retains the $q\bar q$ structure for a meson. The relevant form
factors can be extracted by choosing the plus component of the
matrix elements in the LFQM. In fact, there is spurious contribution
proportional to the lightlike four-vector $\omega=(2, 0, 0_\perp)$
in transforming the covariant Feymann integral into the light-front
form, which makes the theory non-covariant. The covariance requires
inclusion of the zero-mode effect which eliminates the undesired
$\omega$ dependence. Such development is elaborated in
Ref.~\cite{Jaus1}. In this way, all the form factors that are
necessary to represent the Lorentz structure of a hadronic matrix
element can be calculated on the same footing, which is not possible
in the standard LFQM. In the framework of CLFQM, the vertex function
of a meson (bound state) coupling to its constituent quarks consists
of the momentum part and also the spin part, where the former
describes the momentum distribution of the constituent quarks, and
the latter is constructed from the light-front helicity state
involving the Melosh transformation. In the vertex wave function,
there is a free parameter $\beta$, that will be fixed by the decay
constant of the meson. The fermion line is just represented by the
relativistic propagator. The electroweak vertex is given by the
Standard Model. Following the line of Ref.~\cite{Jaus1}, Cheng, Chua
and Hwang have systematically studied the decay constants and form
factors for the $S$- and $P$-wave mesons in 2003 \cite{Chua2003},
while an update was done in Ref.~\cite{Verma} in two points: i) The
experimental information of the branching fractions wherever
available or the lattice results for the decay constants was used to
constrain the parameters $\beta$ in the wave functions; ii) the
extension to the counterpart with $s$ quark ($D_s$ and $B_s$) has
also been considered.

All the form factors for $B (B_s) \to (P, V, A, S)$ transitions
considered by us have been calculated in
Refs.~\cite{Chua2003,Verma}, and thus we omit to repeat these
calculations. However, we note that to make a direct comparison
between theory and experiment, we should further provide the
branching fractions which are the true observables in experiment and
can be directly accessed to test our theoretical predictions. This
constitutes one of our main results. Similar studies have been done
for the case of charmed meson, $D$ and $D_s$ decay \cite{KangSemi},
where the formalism corresponding to the differential decay rates
and branching fractions are explicitly given. Those expressions are
certainly applicable to the $B$ and $B_s$ decay with only some
replacements of the relevant masses. While Ref.~\cite{KangSemi} has
aroused great interest of BES colleagues and some of our results
have been confirmed, a natural question is what will happen in the
beauty $B$ and $B_s$ cases. Combining the running Belle-II, the
predictions for the branching fractions of various channels are
important for our experimental colleagues. The future measurements
will provide valuable information on the form factors as well as the
structure of the axial-vector mesons and scalar mesons, as already
mentioned in the Introduction.

Here we discuss the difference and merit of measuring $B$ meson
decays comparing to $D$ decays. The mass of B meson is heavy enough
that the methods of Perturbative QCD and Soft-Collinear Effective
Theory are available, while there are little reliable theoretical
tools to treat the corresponding $D$ decays. Therefore, some $B$ and
$B_s$ decay channels considered by us in the manuscript have also
been calculated in such approaches, as shown in e.g.,
Ref.~\cite{WeiWang}. We compared our results with them and a nice
agreement is achieved. Consequently, we show the experimental values
or the ones reported by PDG in the tables. On the other hand, the
$D$ meson has smaller phase space resulting in small branching
fractions for decaying into $f_0(1310),\,f_0(1500),\,f_0(1710)$,
e.g., $\mathcal{B}(D^+\to f_0(1710) e^+ \nu_e)\sim 10^{-9}$
\cite{KangSemi} can not be measured at the BESIII factory due to
limited statistics, but for the $B$ meson case, the corresponding
branching fraction is at the order of $10^{-6}$ which is in the
scope of Belle and Belle-II detectors.

Besides the larger phase space available for the $B$ decay, the
experimental situation is also much better. There are much more $B$
data samples than the ones of $D$, and especially when considering
the running of the Belle-II, as the upgrade of Belle detectors. This
constituents one of our direct motivations to reconsider the $B$ and
$B_s$ decay. The total event numbers of $D/D_s$ and $B/B_s$ pairs
are listed in Tabs.~\ref{tab:NBES} and ~\ref{tab:NBelle},
respectively. Clearly, we can see the difference of the order in
magnitude. For example, the branching fractions $\mathcal{B}(D^+\to
f_0(1500)e^+\nu_e)\sim 1\times 10^{-6}$ and $\mathcal{B}(D^+\to
f_0(1710)e^+\nu_e)\sim 5\times 10^{-9}$ \cite{KangSemi} are hard to
be detected by BESIII collaboration, while Belle-II is capable of
measuring those channels due to $\mathcal{B}(B\to
f_0(1500)\ell\nu_\ell)\sim 8\times 10^{-6}$ and $\mathcal{B}(B^+\to
f_0(1710)\ell\nu_\ell)\sim 2\times 10^{-6}$.

\begin{table}[htbp]
 \centering
\begin{tabular*}{0.9\linewidth}{@{\extracolsep{\fill}}lllll}
 \hline \hline
                &current   & planned \\
 \hline\hline
$D^+D^-$        &$(8.296\pm0.031\pm0.064)\times 10^6$  & $\sim 5\times 10^7$\\
 $D^0 \bar D^0$ &$(10.597\pm0.028\pm0.087)\times 10^6$ & $\sim 6.4\times 10^7$\\
 $D_s\bar D_s$  &$\sim 3.3\times 10^6$                  & $\sim 2\times10^7$ \\
 \hline \hline
\end{tabular*}
\caption{The total numbers of $D^+D^-,\,D^0\bar D^0,\, D_s^+D_s^-$
pairs from BESIII collaboration, where in the data-taking plan the
future data samples will be 6 times as large as the current ones.
The number of $D\bar D$ pair is from
Ref.~\cite{1802}.}\label{tab:NBES}
\end{table}

\begin{table}[htbp]
 \centering
\begin{tabular*}{0.9\linewidth}{@{\extracolsep{\fill}}lllll}
 \hline \hline
                &Belle   &BelleII \\
 \hline\hline
$B\bar B$        &$(7.72\pm0.11)\times 10^8$  & $\sim 3.9\times 10^{10}$\\
$B_s\bar B_s$    &$(6.53\pm0.66)\times 10^6$  & $\sim 3.3\times10^8$ \\
 \hline \hline
\end{tabular*}
\caption{The total numbers of $B\bar B$ and $B_s^+ B_s^-$ pairs from
Belle collaboration, while BelleII will have the data samples of 50
times as large as Belle by the mid of next decade. The number of
$B\bar B$ and $B_s\bar B_s$ pairs for Belle collaboration are from
Refs.~\cite{Belle-BBbar,Belle-BsBsbar}.} \label{tab:NBelle}
\end{table}

We finally make two supplemental remarks: (1) in
Refs.~\cite{Chua2003, Verma} all the form factors correspond to the
$V-A$ current, and the ``x'' in Fig.1b therein \cite{Chua2003,Verma}
denotes the insertion of $W$ boson, thus we will not consider the
$b\to s$ transition, penguin operators etc. Or stated differently,
transitions like $B\to K, K^*, K_1(1270), K_1(1400)$ or $B_s\to
\phi$ will not be considered in the current framework. (2) In
general, for the transitions involving the (axial-) vector meson,
the form factors receive the contribution of additional $B$
functions, and for our current case, those have been confirmed to be
negligibly small by numerical calculations \cite{Chua2003}. The
issues concerning self-consistency and covariance of the light-front
quark models are discussed in Ref.~\cite{QinChang}.

\section{Results and discussion}
In this part, we report our theoretically predicted branching
fractions for various semileptonic decay channels considered by us,
and discuss some details and implications.

In Tab.~\ref{tab:meson list}, we list out the mesons that we
considered in the $B$ and $B_s$ semiletonic decays:
\begin{itemize}
\item $P$ denotes the pseudoscalar bosons containing
$\pi,\,K,\,\eta,\,\eta',\,D^0,\,D_s$, \item $V$ contains the vector
mesons $\rho,\,\omega,\,\phi,\,K^*,\,D^*,\,D_s^*$.
\end{itemize}

$S$ contains the heavy scalar nonet
$a_0(1450),\,f_0(1370),\,f_0(1500)/f_0(1710), K^*(1430)$ suggested
by $q\bar q$ quark model \cite{Close}, and the charmed mesons
$D^*_0(2400)$, $D_{s0}^*(2317)$, i.e., the calculation and results
are based on assuming them as the conventional $q\bar q$ meson. In
fact, except for the structures of $a_0(1450)$ and $K^*_0(1430)$
which are less controversial, those of others still need to be
ascertained. Especially, a common viewpoint is to interpret
$D_{s0}^*(2317)$ as a $DK$ molecular or a tetraquark state, see
recent reviews in Refs.~\cite{review1,review2,review3,review4}. The
$D^*_0(2400)$ \footnote{The near mass degeneracy of $D^*_0(2400)$
and $D^*_{s0}(2317)$ is explained by the hadronic loop effects using
the heavy meson chiral perturbation theory \cite{ChengHMChPT}.} is
the excited state of $D$ meson and can be understood from the
heavy-quark spin symmetry, where the light system has $j=s_q + L$,
with $s_q$ denoting the spin of light quark and $L$ the orbital
angular momentum, and thus there are two doublets with $J^P$ as:
\begin{eqnarray} \label{eq:Pcharm1}
j=1/2, \qquad \begin{pmatrix} D^*_0=D^*_0(2400)& \quad 0^+
\\ D'_1=D'_1(2430) & \quad 1^+\end{pmatrix}
\end{eqnarray}
and
\begin{eqnarray}\label{eq:Pcharm2}
j=3/2, \qquad \begin{pmatrix} D_1=D_1(2420)& \quad 1^+
\\ D^*_2=D^*_2(2460) & \quad 2^+\end{pmatrix}.
\end{eqnarray}
However, other interpretations also exist, e.g., $D^*_0(2400)$ shows
two-pole structure and the lower pole associated with
$D^{*}_{s0}(2317)$ forms an SU(3) multiplet \cite{D*2400-1}.
Concerning the $f_0$ states, it is generally argued that they are
the $q\bar q$ meson mixed by glueball contents, but differing in
which state is dominated by glueball component in literature
\footnote{Again, there are also other interpretations, e.g.,
$f_0(1370)$ and $f_0(1710)$ as bound state of two vector mesons
\cite{Osetf0(1370),Osetf0(1710)}.}. Considering the available
lattice and experimental information, the authors of
Refs.~\cite{Chengf01,Chengf02} have done a careful analysis: under
the assumption of the exact SU(3) symmetry, $f_0(1500)$ is an SU(3)
isosinglet octet state and is degenerate with $a_0(1450)$; in the
absence of glueball-quarkonium mixing, $f_0(1710)$ would be a pure
glueball and $f_0(1370)$ a pure SU(3) singlet; when the
glueball-quarkonium mixing is turned on, there will be additional
mixing between the glueball and the SU(3)-singlet, and then
\begin{eqnarray}\label{eq:f0mixing}
\begin{pmatrix} f_0(1370) \\ f_0(1500) \\
f_0(1710)\end{pmatrix}=\begin{pmatrix} 0.78(2) & 0.52(3) & -0.36(1)
\\ -0.55(3) & 0.84(2) & 0.03(2) \\  0.31(1) & 0.17(1) & 0.934(4)  \end{pmatrix}
\begin{pmatrix}f_{0q} \\ f_{0s} \\G \end{pmatrix}
\end{eqnarray}
where the number in the parenthesis indicates the uncertainty for
the last digit of the central value; the scalar $f_{0q}$ ($f_{0s}$)
is the pure $q\bar q$ ($s\bar s$) states with the spin-parity
$J^P=0^+$, whose mass is 1.474 GeV (1.5 GeV), while the glueball
($G$) is 1.7 GeV \cite{Chengf01,Chengf02}. Clearly, $f_0(1710)$
contains mainly glueball and $f_0(1500)$ has the flavor octet
structure. To be specific, we show the corresponding $B\to f_{0q}$
and $B_s\to f_{0s}$ transition form factors \cite{Verma} in
Tab.~\ref{tab:FFf0}.

\begin{table}[htbp]
 \centering
\begin{tabular*}{0.9\linewidth}{@{\extracolsep{\fill}}llll}
 \hline \hline
   $F$             &$F(0)$    & $a$  & $b$\\
 \hline\hline
$F_1^{Bf_{0q}}$       &$0.25 \pm 0.03$  & $1.53 \pm 0.04$ & $0.64^{+0.12}_{-0.09}$\\
$F_0^{Bf_{0q}}$       &$0.25 \pm 0.03$ &$0.54\pm0.07$ & $0.01^{+0.02}_{-0.01}$ \\
$F_1^{B_sf_{0s}}$       &$0.28\pm0.01$  & $1.64\pm0.04$  &$1.07\pm0.12$\\
$F_0^{B_sf_{0s}}$       &$0.28\pm0.01$  & $0.52\pm0.04$  &$0.20\pm0.03$\\
 \hline \hline
\end{tabular*}
\caption{The form factors for $B\to f_{0q}$ and $B_s\to f_{0s}$
transitions \cite{Verma}. }\label{tab:FFf0}
\end{table}

In fact, we wish to stress that the proposed measurements of the
semileptonic $B (B_s)$ decays to $f_0$ states will be a powerful
test for their inner structure due to the absence of the final-state
interaction between $f_0$ and the lepton pair.

The axial-vector mesons $A$ has two kinds of  $1^{++}$ and $1^{+-}$,
where the former contains $a_1(1260),\,f_1(1285),\,f_1(1420)$ and
the latter contains $b_1(1235),\,h_1(1170),\,h_1(1380)$, while
$K_1(1270)$ and $K_1(1400)$ (with $d\bar s$ quark components) mix
since they do not have the definite $C$-parity.

We should consider mixing angle:
\begin{eqnarray}
\eta&=&\eta_q \cos\phi - \eta_s\sin\phi,\nonumber\\
\eta'&=&\eta_q\sin\phi + \eta_s\cos\phi,
\end{eqnarray}
with $\eta_q=(u\bar u+ d\bar d)/\sqrt{2}$ and $\eta_s=s\bar s$, and
$\phi=39.3^\circ\pm1.0^\circ$ extracted from Refs.~\cite{Kroll} is
consistent with the recent result $\phi=42^\circ\pm2.8^\circ$ from
the analysis of the CLEO data \cite{Hietala}.

Let $A_L$ be the light axial vector, and $A_H$ the heavier one.
\begin{eqnarray}
A_L&=&\sin\alpha_A A_q +\cos\alpha_A A_s,\nonumber\\ A_H&=&
\cos\alpha_A A_q - \sin\alpha_A A_s,
\end{eqnarray}
where $A_q$ and $A_s$ denotes the corresponding components $(u\bar
u+ d\bar d)/\sqrt{2}$ (note the factor of 1/2 for calculating the
branching fraction) and $s\bar s$ in the wave functions. Following
the strategy in Refs.~\cite{KangSemi,ChengMixingAngle} we will take
the values $\alpha_{f_1}=69.7^\circ,\,\alpha_{h_1}=86.7^\circ$.
Recently, $h_1(1380)$ has been confirmed by the BES-III
collaboration \cite{BES1380} in the decay channel $J/\psi\to \eta'
K\bar K \pi$, where its mass and width, and the product branching
fraction have been measured. Also, the mixing angle is determined to
be $90.6^\circ\pm2.6^\circ$ \cite{BES1380} based on the mixing angle
$\theta_{K_1}=34^\circ$ and the masses of the axial-vector mesons.
This is consistent with the value that is adopted by us above.
Clearly, the quark contents of $h_1(1170)$ is dominated by $h_{1q}$,
while the $h_1(1380)$ mainly consists of $s\bar s$. Note that in the
literature, e.g., Ref.~\cite{DaiLY}, the mixing angle $\theta$ is
often referred to the singlet-octet one, and $\alpha = \theta +
54.7^\circ$. An ideal mixing is defined as $\tan \theta =
1/\sqrt{2}$, i.e., $\theta= 35.3^\circ$.

The physical mass eigenstates $K_1(1270)$ and $K_1(1400)$ are the
mixture of the $^1P_1$ state $K_{1B}$ and $^3P_1$ state $K_{1A}$
\cite{PDGreview},
\begin{eqnarray}\label{eq:K1mixing}
K_1(1270)&=&K_{1A}\sin\theta_{K_1}+K_{1B}\cos\theta_{K_1},\nl
K_1(1400)&=&K_{1A}\cos\theta_{K_1}-K_{1B}\sin\theta_{K_1},
\end{eqnarray}
and we will take $\theta_{K_1}=33^\circ$ from the analysis of
Ref.~\cite{ChengMixingAngle}.

As mentioned in Sec.~II, the form factors and the formula for
calculating the branching fractions can be found in
Refs.~\cite{Chua2003,Verma} and ~\cite{KangSemi}, respectively. Only
one point is needed to be notified: generally, the form factor is
expressed by
\begin{eqnarray}\label{eq:Fq}
F(q^2)=\frac{F(0)}{1-a (q^2/m_B^2) +b (q^2/m_B^2)^2},
\end{eqnarray}
with the parameters $F(0), a, b$ given in Ref.~\cite{Verma}. As
discussed in \cite{Chua2003}, the form factor $V_2(q^2)$ for
$B(B_s)\to A (1^{+-})$ transition approaches zero at very large
$-|q^2|$ where the three-parameter parametrization,
Eq.~\eqref{eq:Fq}, becomes questionable. Instead, a variant has been
exploited,
\begin{eqnarray}\label{eq:modified Fq}
V_2(q^2)=\frac{V_2(0)}{(1-q^2/m_B^2)[1-a (q^2/m_B^2) +b
(q^2/m_B^2)^2]}.
\end{eqnarray}
The form factor with the expression of Eq.~\eqref{eq:modified Fq} is
applied to the $^1 P_1$ case, i.e., $b_1,\,h_1$ and $K_{1B}$. One
may consider to replace $m_B$ by $m_{B_s}$ in Eqs.~\eqref{eq:Fq} and
\eqref{eq:modified Fq} for $B_s$ decays, however, such difference is
negligible, in practice. We are now in position to provide the
values for the branching fractions, which are listed in
Tabs.~\ref{tab:Be} and ~\ref{tab:Btau} for $B$ decays and
Tab.~\ref{tab:Bs} for $B_s$ decays. Generally, the small masses of
the electron and muon compared to the one for $B$ meson does not
make visible difference for the corresponding branching fractions,
as also stated in PDG ``$\ell$ denotes $e$ or $\mu$''.

\begin{figure}
\begin{center}
\includegraphics[width=85mm]{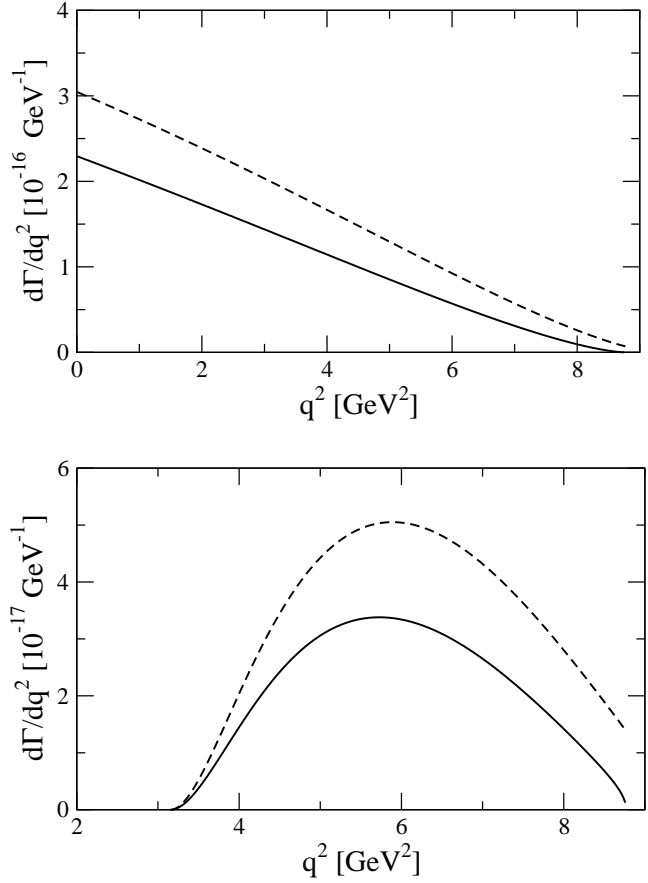} \caption{The differential decay rate
$d\Gamma/dq^2$ as a function of momentum transfer $q^2$. In the
upper plot, the solid and dashed line indicate $B^+\to D^*_0(2400)
e^+ \nu_e$ and $B_s\to D^{*-}_{s0}(2317) e^+ \nu_e$, respectively,
while the lower one shows the corresponding tau lepton modes. }
\label{fig:dGdq2}
\end{center}
\end{figure}

Several remarks are in order:

\begin{itemize}

\item All the $B^+\to P (V) e^+ \nu_e$ modes have been measured by
experiment and our values agree very well with the values reported
by PDG within one standard deviation around. Certainly, our results
may even better match some specific measurements, e.g.,
$\mathcal{B}(B^+\to \bar D^0 e^+ \nu_e)=(2.29\pm0.08\pm0.09)\%$ by
the BaBar collaboration \cite{BaBarBD0}, $\mathcal{B}(B^+\to \bar
D^{*0} e^+ \nu_e)=(6.50\pm0.20\pm0.43)\%$ by CLEO \cite{CLEOBD*0},
and $\mathcal{B}(B^+\to \omega e^+ \nu_e)=(1.35\pm0.21\pm0.11)\times
10^{-4}$ by BaBar \cite{BaBarBomega}. The experimental results are
not yet available for $B^+\to S (A) e^+ \nu_e$ modes. The branching
fractions for semileptonic $B^+\to a_0(1450), a_1(1260), b_1(1235),
f_1(1285), h_1(1170)$ transitions are predicted to be at the order
of $10^{-5}$, while those of other transitions are at the order of
$10^{-6}$. Considering the secondary decays $\mathcal{B}
(a_0(1450)\to \pi\eta)=0.093\pm0.020$, $\mathcal{B} (a_0(1450)\to
\pi\eta')=0.033\pm0.017$ and $\mathcal{B} (a_0(1450)\to K\bar
K)=0.082\pm0.028$, and $\mathcal{B}(f_1(1285)\to
4\pi)=(33.5^{+2.0}_{-1.8})\%$ \cite{PDG}, the statistics for Belle
and Belle-II should be enough for measuring the transition $B^+\to
a_0(1450), \, f_1(1285) e^+\nu_e$. The precise determination of the
pole position (mass and width ) for $f_0(1370)$ is still
challenging. PDG \cite{PDG} shows that its pole position is at
$(1200 - 1500) - i (150 - 250)$ MeV, and the Breit-Wigner or
K-matrix mass and width at $(1200 - 1500)- i (200 - 500)$ MeV,
suffering from large uncertainty. We thus refrain from showing the
branching fractions involving $f_0(1370)$.

\item BaBar \cite{BaBarD*0(2400)} and Belle \cite{BelleD*0(2400)} have
measured the four semileptonic decay modes involving the P-wave
charmed mesons, cf.~Eqs.~\eqref{eq:Pcharm1} and ~\eqref{eq:Pcharm2},
which of course, includes $D^*_0(2400)$. The PDG average value of
$(2.5\pm0.5)\times 10^{-3}$ means the joint branching fraction
$\mathcal{B}(B^+\to \bar D^*_0(2400)\ell\nu_\ell,\,\bar D^*_0(2400)
\to D^-\pi^+)$. Comparing with our theoretical value for $B^+\to
D^*_0(2400)e^+\nu_e$ in Tab.~\ref{tab:Be}, we expect the mode of
$D^*_0(2400)\to D\pi$ is the dominant one in $D^*_0(2400)$ decays.
In fact, the semileptonic decay $B\to D^*_0(2400)$, as a background
contributing to one of the leading sources of the systematical
uncertainty for the extraction of $|V_{cb}|$ from $B\to D^*
\ell\nu_\ell$, is still poorly known, see the review in
Ref.~\cite{Semi2016}. The Belle-II and LHCb detectors will provide
the opportunity for the precision measurements. We also display the
differential decay rate for $B\to D^*_0(2400)$ as well as $B\to
D^*_{s0}(2317)$ + lepton pairs  in Fig.~\ref{fig:dGdq2} for
convenience of comparison with the future experiments.

\item Due to the factor of $|V_{cb}/V_{ub}|^2\approx 115$, the
branching fraction of $b\to c$ decay is generally enhanced by two
orders compared to $b\to u$ decay, as can be seen in
Tabs.~\ref{tab:Be},~\ref{tab:Btau} and ~\ref{tab:Bs} (Note that the
additional factor of $\sin^2\phi$ or $\cos^2\phi$ appears in the
processes of $B^+\to\eta, \,\eta'$ to calculate the branching
fractions). $\mathcal{B}(B_s\to D_s^- + X)=(93\pm25)\%$ \cite{PDG}
again shows the dominance of $b\to c$ tansition. In
Tab.~\ref{tab:Bs}, the $B_s$ decay branching fraction is at the
order of $10^{-4}$ for $b\to u$ and $10^{-2}$ for $b\to c$.
Unfortunately, there is scarce experimental information on the
semileptonic $B_s$ decay except for the inclusive semileptonic decay
$\mathcal{B}(B_s\to X \ell\nu_\ell)=(9.6\pm0.8)\%$ \cite{PDG}. As
can be clearly seen, the sum of the branching fractions for the
channels considered in Tab.~\ref{tab:Bs} does not exceed this limit.
The theoretical predictions for $\mathcal{B}(B_s\to D_s \ell
\nu_\ell)$ vary from 1.0\% to 3.2\% and for $\mathcal{B}(B_s\to
D_s^* \ell \nu_\ell)$ vary from 4.3\% to 7.6\% \cite{BelleBs}, see
e.g., Refs.~\cite{BsSemi1,BsSemi2,BsSemi3,BsSemi4};
$\mathcal{B}(B_s\to D_{s0}^*(2317)\ell\nu_\ell)\sim 0.20\% - 0.57\%$
\cite{BsSemi4,LuCD,LiuX}. Regarding $D^*_{s0}(2317)$ as a $DK$
molecular state, the authors of Ref.~\cite{Oset2317} predict
$\mathcal{B}(B_s\to D_{s0}^*(2317)\ell\nu_\ell) = 0.13\%$. The
process $B_s\to K^- \ell\nu_\ell$ has been calculated in
Refs.~\cite{BsSemi4,Nieves} and also examined in lattice QCD
\cite{lattice1,lattice2}. The $B_s\to K_J^* \ell\nu_\ell$ decay is
investigated in Ref.~\cite{GalkinKJ}, and our result for $B_s\to
K^{*-}_0(1430)\ell\nu_\ell$ agrees very well with theirs, but not
for $K_1(1270)$ and $K_1(1400)$ sector. Adopting their values of the
mixing angles as input still does not remedy such discrepancy, so we
will regard such discrepancy as the different predictions from the
two models. Given the branching fractions of
$\mathcal{B}(K_1(1270)\to K\rho)=(42\pm6)\%$ and
$\mathcal{B}(K_1(1400)\to K^*(892)\pi)=(94\pm6)\%$, the $B_s\to K_1$
transitions could be measured with the current statistics in
Belle/Belle-II and LHCb. Overall speaking, little has been known for
the experimental information on the exclusive semileptonic $B_s$
decay, while plentiful theoretical predictions have been done. This
situation highly calls for the true experimental measurements, and
this can be realized with Belle/Belle-II and LHCb detectors.

\item We wish to comment that even-parity light mesons, including the
axial-vector meson, the scalar meson above 1 GeV, and the $P-$wave
charmed meson, can be also studied via hadronic two-body $B$ decays
within the factorization scheme
\cite{ChengTwobody1,ChengTwobody2,ChengTwobody3,ChengTwobody4}. The
semileptonic decay modes investigated here will provide a much
cleaner environment to explore the nature of these mesons owing to
the absence of the strong hadronic final-state interactions
manifested in the two-body hadronic decay. At least, the
investigation of such semileptonic modes could serve as a supplement
to the hadronic two-body decay.

\item The CKM matrix element $|V_{ub}|$ suffers from large uncertainty around 19\%,
while $|V_{cb}|$ has been determined better with the uncertainty of
$4\%$. Roughly assigning 10\% error induced by form factors, we have
the combined uncertainty of 22\% and 11\% for the processes $b\to u$
and $b \to c$, respectively. Additionally, the uncertainty induced
by the mixing angle needs more care. Guided by
Ref.~\cite{ChengMixingAngle} we allow the variations of
$\alpha_{f_1},\,\alpha_{h_1},\,\theta_{K_1}$ within
$8^\circ,\,6^\circ,\,4^\circ$, in order \footnote{These
uncertainties are also used in Ref.~\cite{KangSemi}. We notice that
the error for $\alpha_{h_1}$ agrees very well with the very recent
determination of $7.2^\circ$ by the BESIII collaboration
\cite{BES1380}.}, which may produce the (very) asymmetry error. In
such cases, we show in the brackets the resulting allowed regions
for the branching fractions. The branching fractions for $K_1$ case
is not sensitive to the mixing angle $\theta_{K_1}$. The mixing
angle $\alpha_{h_1}$ for $h_1(1170)$ and $h_1(1380)$ states crosses
$90^\circ$, where $h_1(1170)$ purely consists of $q\bar q$ and
$h_1(1380)$ purely $s\bar s$. This shows the origin of vanishing
branching fractions of $B\to h_1(1380)\ell\nu_\ell$ in
Tabs.~\ref{tab:Be} and ~\ref{tab:Btau}. In some cases, the branching
fractions are sensitive to the mixing angles, e.g., for $B\to
f_1,\,h_1$. From this point of view, it should be understood that
the future measurements on these channels will be highly meaningful
for a ``precise'' determination of the mixing angles, as also
mentioned in the Introduction.

\item There are several recent tests of the lepton
universality via the semeileptonic decay modes, e.g.,
$\mathcal{B}(D\to \pi \mu \nu_\mu)/\mathcal{B}(D\to \pi e \nu_e)$
done by the BES-III collaboration \cite{BESLFU}, and
$\mathcal{B}(B\to D^* \tau \nu_\tau)/\mathcal{B}(B\to D^* \mu
\nu_\mu)$ done by the LHCb collaboration \cite{LHCbLFU}. Motivated
by this, we also calculate the branching fractions of the
semileptonic decays involving the $\tau$ lepton mode. Our results
agree very well with the experimental values $\mathcal{B}(B^+\to
\bar D^0\tau\nu_\tau)=(7.7\pm2.5)\times 10^{-3}$ and $B^+\to\bar
D^{*0}\tau\nu_\tau=(1.88\pm0.20)\%$ \cite{PDG}. That is, what we
compared is the direct number of the branching fraction of the
$\tau$ mode, but not the ratio between the $\tau$ and electron one.
The latter is related to the recently well-known $R_D$ or $R_{D^*}$
puzzle \cite{HFLAG}. In fact, we do not touch this issue since we
are working in the framework of Standard Model (keeping the lepton
universality) and also there is the tricky estimate of
uncertainties. However, as a passing comment, we want to remind the
importance of the precise determination of the uncertainties.
Starting from the decay $B\to D^* \ell \bar \nu_\ell$, the authors
of Ref.~\cite{Dutta} also discussed the corresponding strange quark
partner, $B_s\to D_s \tau \nu_\tau$, which is also investigated by
us. Very recently, $B^*\to (D,\,D_s,\,\pi,\,K) \ell\nu_\ell$ is also
discussed for probing the New Physics effects \cite{Changqin}. On
the experimental aspect, the electron mode is usually the easiest
one to be measured, while one may encounter the large
misidentification between $\mu$ and $\pi$ \footnote{The decaying of
muon to electron occurs outside the detector and thus muon can be
regarded as a stable particle inside the detector.}. For the $\tau$
case, the experimental error will be even larger: the two largest
decay channels of $\tau$ are \cite{PDG} $\mathcal {B} (\tau \to
\mu^- \bar \nu_\mu \nu_\tau) =(17.39\pm0.04)\%$ and $\mathcal {B}
(\tau \to e^- \bar \nu_e \nu_\tau) =(17.82\pm0.04)\%$; both of them
contain two neutrinos, which hinders the full construction resulting
in large background, and also there is no way to use the recoiling
information due to the existence of multi-neutrinos.
\end{itemize}

\begin{widetext}
\begin{center}
\begin{table}[htbp]
\begin{tabular*}{0.9\linewidth}{@{\extracolsep{\fill}}c|lllllll}
\hline \hline
P  &$ \pi$ &$ K$ &$ \eta$ &$\eta'$ &$D^0$ &$D_s$  & \\
\hline
V &$\rho$ &$\omega$  &$K^*$ &$ \phi$ &$D^*$ &$D^*_s$ & \\
\hline
S & $a_0(1450)$ &$f_0(1370)$ &$ f_0(1500)$ &$ f_0(1710)$ &$K^*_0(1430)$ &$D^*_0(2400)$ &$D^*_{s0}(2317)$ \\
\hline \multirow{2}{*}{A} &$1^{++}$ &$a_1(1260)$ &$f_1(1285)$ &$f_1(1420)$ &\multirow{2}{*}{$K_1(1270)$}
&\multirow{2}{*}{$K_1(1400)$} &  \\
 {} &$1^{+-}$ &$b_1(1235)$ &$h_1(1170)$ &$h_1(1380)$ &   & & \\
\hline \hline
\end{tabular*}
\caption{The involved mesons in the semileptonic $B$ and $B_s$
decays, where $P,\,V,\,S,\,A$ denotes the pseudoscalar, vector,
scalar (mass above 1 GeV), and axial-vector mesons, respectively. In
the axial-vector mesons $K_1(1270)$ and $K_1(1400)$ have no definite
$C$-parity and do mix with each other.} \label{tab:meson list}
\end{table}
\end{center}
\end{widetext}

\begin{widetext}
\begin{center}
 \begin{table}[htbp]
\begin{tabular*}{0.9\linewidth}{@{\extracolsep{\fill}}lllll}
 \hline \hline
 Channel ($10^{-5}$) &$B\to \pi^0$   &$B\to \eta$  &$B\to\eta'$  &$B\to \bar D^0$  \\
 Theory     &$7.66\pm1.69$   &$5.27\pm1.16$ &$2.56\pm0.56$  &$2608\pm287$  \\
 PDG  &$7.80\pm0.27$ &$3.9\pm0.5$ &$2.3\pm0.8$ &$2200\pm100$ \\
 \hline
 Channel ($10^{-4}$) &$B\to \rho^0$ &$B\to \omega$  &$B\to \bar D^{*0}$ &  \\
 Theory   &$2.13\pm0.47$  &$2.0\pm0.44$   &$671\pm74$  & \\
 PDG      &$1.58\pm 0.11$ &$1.19\pm0.09$ &$488\pm10$ & \\
 \hline Channel ($10^{-5}$) &$B\to a_0(1450)$ &$B\to f_0(1500)$ &$B\to f_0(1710)$ &$B\to \bar D^*_0$ \\
 Theory  &$2.72\pm0.60$  &$0.77\pm0.17$  &$0.21\pm0.05$   &$231\pm25$ \\
 PDG  & & & &$250\pm 50$   \\
\hline Channel ($10^{-5}$) &$B\to a_1(1260)$ &$B\to f_1(1285)$ &$B\to f_1(1420)$ & \\
Theory  &$6.3\pm1.4$    &$\{3.8,\,7.3\}$   &$\{0.17,\, 1.34\}$ &
\\[1.5ex]
Channel ($10^{-5}$) &$B\to b_1(1235)$ &$B\to h_1(1170)$ &$B\to h_1(1380)$ & \\
Theory &$7.7\pm1.7$ &\{6.7,\, 10.8\}  &$\{0,\, 0.16\}$  & \\
 \hline \hline
\end{tabular*}
\caption{Our theoretical predictions for the branching fractions of
semileptonic $B^+$ decays, $B^+\to (P,\,V,\,S,\,A) e^+\nu_e$,
confronting with the PDG values \cite{PDG} if available. Units are
shown in the parentheses. The branching fraction $(250\pm 50)\times
10^{-5}$ corresponds to the joint decay $B^+\to \bar
D^*_0(2400)\ell\nu_\ell,\,\bar D^*_0(2400) \to D^-\pi^+$. For the
axial-vector mesons, we vary the mixing angle within the errors, and
show the resulting regions in the brackets.} \label{tab:Be}
\end{table}
\end{center}
\end{widetext}

\begin{widetext}
\begin{center}
 \begin{table}[htbp]
\begin{tabular*}{0.9\linewidth}{@{\extracolsep{\fill}}lllll}
 \hline \hline
 Channel ($10^{-5}$) &$B\to \pi^0$   &$B\to \eta$  &$B\to\eta'$  &$B\to \bar D^0$  \\
 Theory &$5.21\pm1.15$ &$3.23\pm0.71$  &$1.36\pm0.30$ &$783\pm86$\\
 PDG  &   &  &  &$770\pm25$ \\
 \hline
 Channel ($10^{-4}$) &$B\to \rho^0$ &$B\to \omega$  &$B\to \bar D^{*0}$ &  \\
 Theory & $1.16\pm0.26$ &$1.07\pm0.24$ &$166.5\pm18.3$ &\\
 PDG & & &$188\pm20$ & \\
 \hline
 Channel ($10^{-5}$) &$B\to a_0(1450)$ &$B\to f_0(1500)$ &$B\to f_0(1710)$ &$B\to \bar D^*_0$ \\
 Theory &$1.04\pm0.23$ &$0.29\pm0.06$ &$0.07\pm0.02$ &$29.7\pm3.3$\\
\hline
Channel ($10^{-5}$) &$B\to a_1(1260)$ &$B\to f_1(1285)$ &$B\to f_1(1420)$ & \\
Theory &$2.68\pm0.59$  &$\{1.56,\, 3.05\}$ &$\{0.06,\,0.52\}$ & \\[1.5ex]
Channel ($10^{-5}$) &$B\to b_1(1235)$ &$B\to h_1(1170)$ &$B\to h_1(1380)$ & \\
Theory &$3.0\pm0.7$ &\{2.57,\, 4.12\} &\{0,\, 0.06\} & \\
 \hline \hline
\end{tabular*}
\caption{Same as Tab.~\ref{tab:Be}, but for the tau lepton modes.}
\label{tab:Btau}
\end{table}
\end{center}
\end{widetext}

\begin{widetext}
\begin{table}[htbp]
\centering
\begin{tabular*}{0.9\linewidth}{@{\extracolsep{\fill}}lll}
\hline \hline
Channel ($10^{-4}$) &$B_s\to K$ &$B_s\to D_s$ \\
Theory   &$1.0\pm0.22$ ($0.68\pm0.15$) &$245\pm27$ ($73.3\pm8.1$)\\
\hline
Channel ($10^{-4}$) &$B_s\to K^*$ &$B_s\to D^*_s$  \\
Theory &$3.3\pm0.73$ ($1.72\pm0.38$) &$605\pm67$ ($151\pm17$) \\
\hline
Channel ($10^{-5}$) &$B_s\to K^*_0(1430)$ &$B_s\to D^{*-}_{s0}$ \\
Theory   &$6.05\pm1.33$ ($2.55\pm0.56$) &$307\pm34$ ($44.4\pm4.9$)  \\
\hline
Channel ($10^{-4}$) &$B_s\to K_1(1270)$  &$B_s\to K_1(1400)$ \\
Theory  &$2.47 \pm 0.57$ ($0.98\pm0.23$) &$0.21\pm 0.05$ ($0.09\pm0.02$)  \\
\hline \hline
\end{tabular*}
\caption{Same as Table \ref{tab:Be} but for the $B_s$ decays, with
the numbers in the parentheses indicate the branching fractions for
the corresponding $\tau$ modes.}
 \label{tab:Bs}
\end{table}
\end{widetext}

\section{Conclusion}
Based on the analysis of the form factors from the covariant
light-front quark model \cite{Chua2003,Verma}, we provide the
branching fractions for $B\to (P,\,V,\,S,\,A) \ell \bar\nu_\ell$
with $P,\,V,\,S,\,A$ denoting the corresponding pseudoscalar,
vector, the scalar mesons with mass above 1 GeV, and the axial-vecor
mesons, respectively. Those mesons are listed in Table
\ref{tab:meson list}. Under the framework of the lepton flavor
universality, the branching fractions for the semileptonic decay
involving the $\tau$ mode are also provided. The predicted branching
fractions are typically in the range of $10^{-6}\sim 10^{-4}$. On
the experimental side, $(772\pm11)\times 10^6$ $B\bar B$ and
$(6.53\pm0.66)\times 10^6$ $B_s\bar B_s$ pairs have already been
collected by the Belle detector, and Belle-II will have a larger
statistics with 40 times more than Belle. Those decay modes can be
accessed by the Belle, Belle-II and LHCb data samples, which renders
the test of theoretical calculation, and more importantly, provides
the valuable information on the structure of the scalar and
axial-vector meson, e.g., the weights of quark-antiquark components
in the $f_0(1370),\,f_0(1500),\,f_0(1710)$ states
(cf.~Eq.~\eqref{eq:f0mixing}), the mixing angles for
$f_1(1285)-f_1(1420)$ and $h_1(1170)-h_1(1380)$ states.

Assuming $D^*_0(2400)$ and $D^*_{s0}(2317)$ as the conventional
quark-antiquark mesons, we predict the branching fractions of
$\mathcal{B}(B\to D^*_0(2400) \ell \nu_\ell)=(2.31\pm 0.25)\times
10^{-3}$ and $\mathcal{B}(B_s\to D^{*-}_{s0}(2317) \ell
\nu_\ell)=(3.07\pm 0.34)\times 10^{-3}$. Confronting these values
with future experimental results will provide a further scrutiny for
the possible assignment of $q\bar q$ interpretation.

\section*{Acknowledgement}
The author XWK acknowledges helpful discussions with
Prof.~R.~C.~Verma. We also thank Prof.~Ulf-G.Mei{\ss}ner for a
careful reading. Prof.~Hai-Yang Cheng is greatly acknowledged for
his suggestions to improve the quality. This work is sponsored by
NSFC Grant Nos. 11805012, 11805037, U1832121, 11805059, 11805153,
11881240256, and also by Shanghai Pujiang Program under Grant
No.18PJ1401000 and Open Research Program of Large Research
Infrastructures (2017), Chinese Academy of Sciences; DFG (SFB/TR
110, ``Symmetries and the Emergence of Structure in QCD''); Natural
Science Basic Research Plan in Shaanxi Province of China (Program
No. 2018JQ1002).

\end{document}